\documentclass[preprint,review,11pt,times]{elsarticle}
\usepackage{graphicx,color,soul,amssymb,bm,framed,multicol,multirow,subfig,array,rotating,threeparttable,makecell,textcomp,stfloats,xcolor,lscape,nomencl,hhline,setspace,mathrsfs,indentfirst,balance,enumitem,boldline,booktabs,colortbl,geometry,amsmath}%
\usepackage[numbers]{natbib}
\usepackage[T1]{fontenc}
\usepackage[colorlinks = true,linkcolor=blue, urlcolor=blue, citecolor=blue, anchorcolor = blue]{hyperref}

\journal{Soft Matter}
\begin{document}
\doublespacing
\begin{frontmatter}

\title{Marine mussel plaque-inspired super soft, ductile and tough lattices}

\author[address 1]{Yong Pang}
\author[address 1]{Tao Liu\corref{Corresponding author}}
\ead{tao.liu@qmul.ac.uk}

\cortext[Corresponding author]{Corresponding author}
\address[address 1]{School of Engineering and Materials Science, Queen Mary University of London, London E1 4NS, UK}

\begin{abstract}
The rise of soft materials and additive manufacturing has provided the feasibility of developing elastomer lattices for various engineering applications. Although earlier attempts have been made to manufacture and test the elastomer lattices, a comprehensive understanding of their fracture behaviour has remained elusive. Inspired by the soft foams in mussel plaque core, the present study introduces the concept of soft, ductile, and tough elastomer lattices. Uniaxial and planar tension tests were conducted, and numerical models were developed to facilitate data interpretation. The applicability of planar tension for measuring the fracture energy of elastomer lattices was confirmed through tensile tests using samples with various crack lengths and double-peeling tests. The proposed elastomer lattices achieved Young’s module of $0.04\sim0.17$ $MPa$, failure strain of $135.5 \sim 213.9\%$, and fracture energy of $582 \sim 941$ $J/m^2$ by manipulating the volume fraction ($20 \sim 50 \%$), positioning them as softer, more ductile, and tougher than the majority of engineering foams in the Ashby diagram. In addition, the fracture energy of the elastomer lattices was sensitive to volume fraction but insensitive to initial crack length, thickness, and height ($\geq20$ $mm$). Notably, two distinct failure modes, truss failure and joint failure, were correlated with volume fraction. The specimens that failed at the transition between these two failure modes achieve a higher fracture energy. This study provides new insights to the materials science community and provides a foundation for the development of lightweight, damage-tolerant, and high-performance metamaterials.

\end{abstract}

\end{frontmatter}

\section{Introduction}
Nature is the main source of inspiration for the design of architected metamaterials. Extensive studies have been conducted to develop high-performance metamaterials by mimicking hierarchical microstructures in a biological system. For example, deep-sea hexactinellid sponges inspired diagonal-braced lattice structure was developed to reinforce buckling resistance \cite{Fernandes2021}. Wood-inspired circular honeycombs \cite{Ha2022}, Glyptotherium osteoderm-inspired switches \cite{Du2018}, and mantis shrimp pincer-inspired laminate \cite{Jia2019} were proposed to enhance energy absorption under impact loads. Bionic bamboo thin-walled structures \cite{Fu2019,Zou2016,Tan2011}, water spiders-inspired reticulated shell structures \cite{Wang2019}, and shellfish nacre-inspired laminate \cite{Gu2017,Dimas2013,Marthe2009} were designed to improve bending, tension, and compression performance.  In the majority of these investigations, bio-inspired metamaterials have been additively manufactured using robust materials such as polycarbonate, aluminum, Ti6Al4V, AlSi10M, and stiff glassy polymer VeroWhite. Although earlier attempts have explored the integration of thin layers of rubber-like materials into the interlayers of nacre-inspired laminates to enhance mechanical performance \cite{Gu2017,Dimas2013,Jia2019}, the primary parts of metamaterials or cellular solids have seldom been fabricated using hyperelastic materials due to inherent manufacturing challenges.

\begin{figure*}[t]
\centering
\includegraphics[width=1 \linewidth]{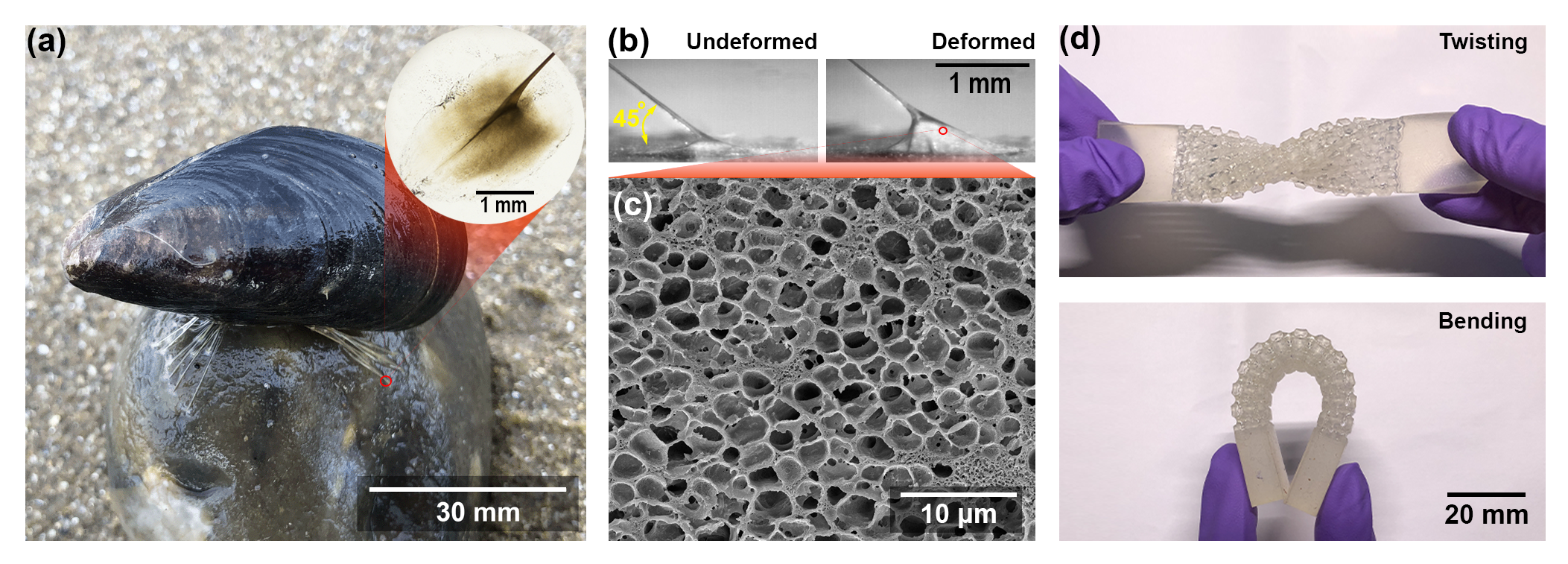}
\caption{(a) A marine mussel anchors to a rock using a plaque-thread system. The inset shows a single mussel plaque. (b) The side view of the un-deformed and deformed mussel plaque. (c) Porous microstructure in a mussel plaque. (d) Mussel plaque-inspired elastomer lattice subjected to twisting and bending.}
\label{Fig1}
\end{figure*}

Recent studies show that marine mussel plaque brings new insights into the development of soft, ductile, and tough cellular solids \cite{Qureshi2022,Cohen2019,Wilhelm2017,Desmond2015}. The marine mussel plaque, a thin biofilm of approximately 100 $\mu m$ in thickness and 2 $mm$ in length, allows mussels to anchor themselves to rocky substrate under wave-swept conditions, as shown in Fig. \ref{Fig1}a. Unlike other stiff cellular structures in nature, marine mussel plaques are a hyperplastic material that undergoes significant deformation under external loads \cite{Pang2023}, as shown in Fig. \ref{Fig1}b. This exceptional deformability is achieved through the collaboration of dense protective cuticle layers \cite{Valois2019,Jehle2020} and porous plaque cores reinforced by collagen fibre bundles \cite{Harrington2010,Filippidi2015,Bernstein2020}. The porous core, a crucial component of mussel plaque (Fig. \ref{Fig1}c), is reminiscent of cellular solids or truss lattices that have superior load-bearing \cite{Fernandes2021,Bhuwal2023,Bhuwal2022}, damage-tolerant \cite{Yang2022,Pham2019}, and energy-absorption capabilities. Therefore, inspired by the marine mussel plaque, it can be hypothesised that hyperplastic materials may benefit the development of engineering lattice or foam structures in terms of failure strain and fracture energy.

To date, a few seminal attempts have been made to manufacture stretchable three-dimensional (3D) lattices using hyperplastic materials. These endeavors fall into two distinct categories. The first group focusses on the creation of stretchable lattices using the direct 3D printing method. For example, Ge et al. \cite{Ge2021} proposed a self-built digital light processing (DLP) 3D printer and proposed the concept of multimaterial lattices consisting of rigid polymer, hydrogel, and elastomer. Luo et al. \cite{Luo2020} and Nicholas et al. \cite{Nicholas2020} developed stretchable lattices using photocurable liquid crystal elastomers,  enhancing energy dissipation under compression. Yan et al. \cite{Yan2020} employed a polyjet 3D printing technique to manufacture multifunctional soft lattices for resistance control, achieving an elongation of up to 230\% under uniaxial tension. The second group fabricates stretchable lattices by casting elastomers into water-soluble hollow scaffolds. Li et al. \cite{Li2017} reported on stretchable 3D lattice conductors made of silicone rubber, exhibiting an elongation of up to 600\% under uniaxial tension. Zhang et al. \cite{Zhang2022} subsequently developed a theoretical model to elucidate the mechanics of stretchy elastomer lattices, explaining their ability to achieve such high elongations. Although the above studies have demonstrated the technical feasibility of manufacturing elastomer lattices and have characterised their uniaxial mechanical behaviour, a gap in current knowledge persists concerning the fracture mechanics of elastomer lattices. Understanding the fracture behaviour of elastomer lattices will provide new insights into the invention of novel, damage-tolerant, multifunctional and lightweight metamaterials.

A well-known pure shear configuration \cite{Rivlin1953} has been extensively used to characterise the fracture energy of soft materials. Sun et al. \cite{Sun2012} applied the pure shear configuration to measure the fracture energy of highly stretchable and tough hydrogels and verified the measurement with a double peeling test. Lee and Pharr \cite{Lee2019} studied the sideways crack propagation in a silicone elastomer using the pure shear configuration. Subsequent studies investigated the fracture and fatigue behaviours of soft materials under cyclic loads \cite{Wang20192, Sanoja2021, Liu2023, Lin2022}. Although the pure shear configuration has been approved as one of the most convenient approaches to measure fracture energy \cite{Shrimali2023,Long2016}, the specific samples used in these studies are incompressible hyperplastic solids, forming a pure shear zone between the crack tip and the force-free edge under tension \cite{Sun2012, Lee2019, Wang20192, Sanoja2021, Liu2023, Lin2022}. However, unlike hyperplastic solids, soft lattice is a compressible material, which means a pure shear zone no longer exists. The applicability of the pure shear configuration for measuring the fracture energy of soft lattices remains uncertain. To the best of our knowledge, there are no generally accepted methods for measuring the fracture energy for soft lattices. Thus, there is an urgent need to re-examine or extend the applicability of current methods for soft lattices.

Inspired by the soft foams in the marine mussel plaque, the present study developed soft, ductile, and tough lattices using ultraviolet-curable elastomers, as shown in Fig. \ref{Fig1}d. First, elastomer lattices with various volume fractions were manufactured. Uniaxial tensile tests and planar tensions were then carried out to investigate the tensile properties and fracture energy of these elastomer lattices, respectively. Subsequently, finite element models were developed accordingly based on these test configurations to assist the data interpretation. In addition, the effects of crack length, sample geometries, and volume fraction on the fracture energy of elastomer lattices were studied and quantified. The outcomes of this study provide new insights into the development and testing methods of high-performance soft lattices.

\section{Experimental methods}
\subsection{Sample preparation}
The elastomer lattices were manufactured by a liquid crystal display (LCD) printer (Phrozen Sonic Mighty 8K) using a clear photo-polymer resin (Liqcreate Elastomer-X). The lattice models were designed using a MATLAB-based package (FLatt Pack Program \cite{Maskery2022}) and sliced using a 3D printing preparation tool (Chitubox). 4 $\times$ 4 $\times$ 4 $mm$ body-centred cubic (BCC) unit cells with various volume fractions (20 $\sim$ 50\%) were employed to manufacture elastomer lattices, as shown in Figs. \ref{Fig2}b and d. Such unit cell size was selected to balance the printing quality and computational costs of FE simulations. To achieve optimal printing quality, various printing parameters were tested. The final printing parameters were selected using a layer thickness of 50 $\mu m$, an exposure time of 26 $s$, a base layer count of 2, a base layer exposure time of 60 $s$, a lift height of 10 $mm$, a lift speed of 60 $mm/min$, and a retract speed of 90 $mm/min$. The manufactured elastomer lattices were rinsed using 85\% ethanol solution in an ultrasonic cleaner for five minutes to wash off all residual unreacted resin and underwent a post-curing process for ten minutes. 

\subsection{Uniaxial tests}
\begin{figure*}[t]
\centering
\includegraphics[width=1 \linewidth]{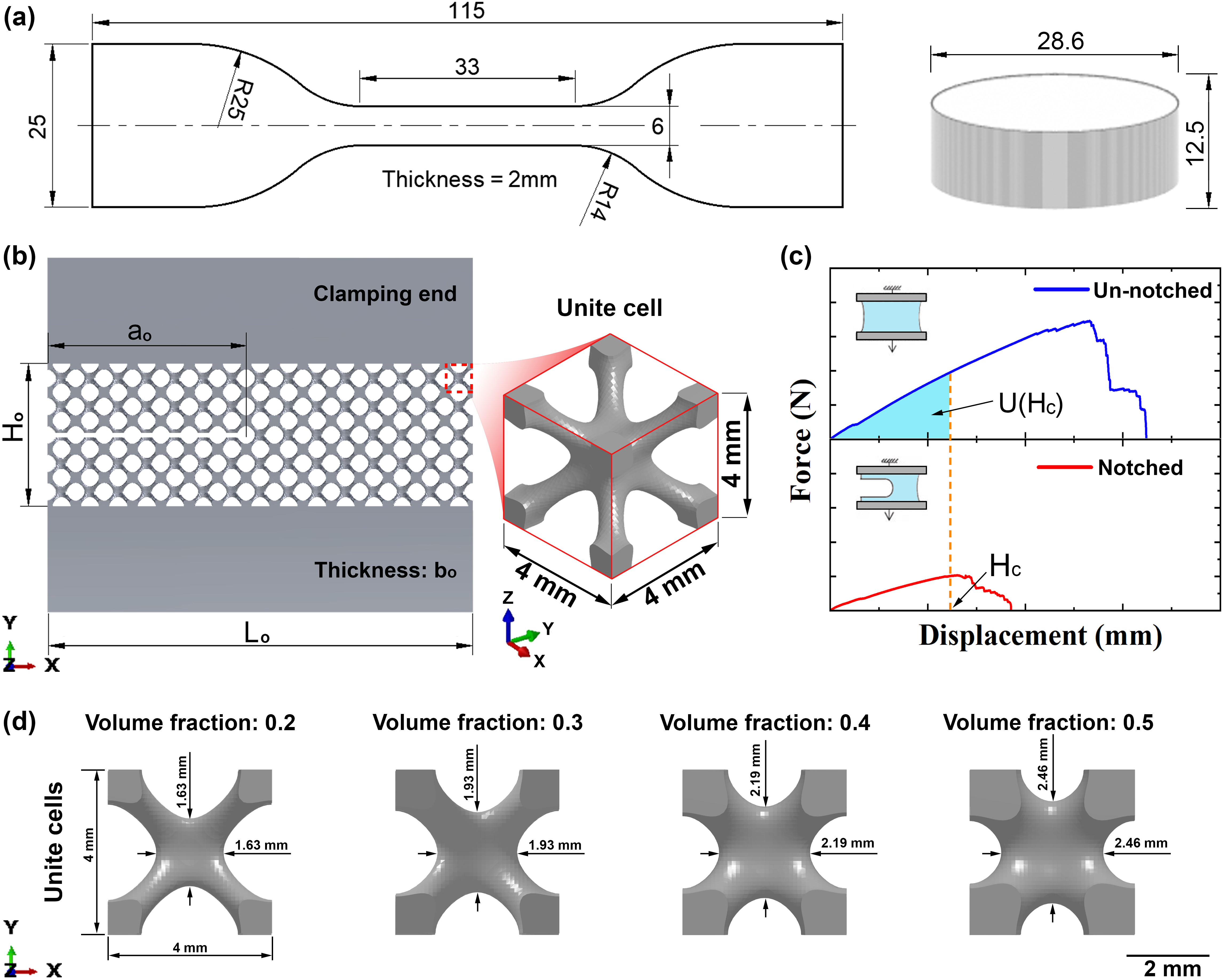}
\caption{(a) Standard uniaxial tension and compression test specimens to determine the mechanical properties of bulk elastomer. (b) Dimensions of notched specimens for planar tension. (c) Determination of fracture energy using notched and un-notched specimens. (d) Dimensions of BCC unit cells with various volume fractions.}
\label{Fig2}
\end{figure*}

Uniaxial tests were conducted on Instron 5900R84 (10 $kN$ load cell) to determine the mechanical properties of bulk material and elastomer lattices. Quasi-static tension and compression were performed using displacement control at a strain rate of 0.001/s. A charge-coupled device (CCD) camera was configured at a special resolution of 100 pixels/mm and frame rate of 0.1 fps to capture the specimen deformation in response to tensile loads. The uniaxial properties of the bulk material were determined using standard tension and compression samples according to ASTM D412 \cite{ASTMD412}, as shown in Fig. \ref{Fig2}a. The tensile properties of the elastomer lattices at various volume fractions ($V_f$ = 0.2 - 0.5) were determined using rectangle tensile specimens in a dimension of $60\times20\times8 mm$ and fitted using \cite{Fleck2010,Gibson2003,Liviu2020}
\begin{equation}
P_l=CP_{s}{V_f}^n 
\label{Eq1}
\end{equation}
where $P_L$ and $P_{s}$ are the mechanical property of the lattice and bulk materials, respectively. They include Young's modulus ($E_s$), tensile strength ($\sigma_s$) and failure strain ($\varepsilon_s$) of the bulk material and Young's modulus ($E_l$), tensile strength ($\sigma_l$) and failure strain ($\varepsilon_l$) of the lattice. $C$ and $n$ are the materials constants. $V_f$ is the volume fraction of the elastomer lattice, defined as
\begin{equation}
V_f=\frac{\rho_l}{\rho_s}
\label{Eq2}
\end{equation}
where $\rho_l$ and $\rho_s$ denote the densities of the elastomer lattice and the bulk materials, respectively. 

\subsection{Planar tension}
Fracture energy of the elastomer lattices was determined under planar tensions using the Instron 5900R84 tester (10 $kN$ load cell). The unnotched specimen used width ($L_o$), height ($H_o$), thickness ($b_o$), and the notched specimens used crack length ($a_o$), as shown in Fig. \ref{Fig2}b.  Various geometries with $H_o$ (4 $\sim$ 44), $b_o$ (4 $\sim$ 24), $a_o$ (8 $\sim$ 48) and $V_f$  (0.2 $\sim$ 0.5) were studied to investigate the effects of dimensions on fracture energy. To achieve pure shear deformation, $H_o$ is required to meet the geometry requirements ($H_o\gg b_o$ and $a_o$) \cite{Anshul2023}. Pure shear tests were conducted at quasi-static condition at a strain rate of 0.001/s. A CCD camera configured under the same conditions as uniaxial tensile tests was used to monitor crack propagation. The fracture energy ($\Gamma$) for a pure shear notched specimen at the start of crack propagation is given by \cite{Sun2012,Rivlin1953} 
\begin{equation}
\Gamma =\frac{U(H_c)}{L_ob_o}
\label{Eq3}
\end{equation}
where, $U(H_c)$ is the work done by the applied force of un-notched specimen, and $H_c$ denotes the critical distance between the clamps of notched specimen, as shown in Fig. \ref{Fig2}c. 

In this paper, tensile strain and stress for notched and un-notched specimens using following convention  
\begin{equation}
\varepsilon_l=\frac{\Delta H_o}{H_o}
\label{Eq4}
\end{equation}
\begin{equation}
\sigma_l=\frac{F}{L_ob_o}
\label{Eq5}
\end{equation}
where, $\Delta H_o$ is the elongation of the samples, and $F$ defined as the reaction force recorded by the load cell. The crack length of the notched specimen was normalised by 
\begin{equation}
\bar{a}=\frac{a_o}{L_o}
\label{Eq6}
\end{equation}

\section{Numerical simulation}
\subsection{Materials constitutive model}
\label{S3.2}
To simulate the mechanical behaviour of the bulk elastomer, the experimental results measured from uniaxial tension and compression were fitted into the three-order Ogden's model \cite{Ogden1972}
\begin{equation}
\Psi (\bar{\lambda}_{1},\bar{\lambda}_{2},\bar{\lambda}_{3})=\sum_{k=1}^{N}\frac{2\mu _{k}}{\alpha _{k}^{2}}(\bar{\lambda}_{1}^{\alpha_{k}}+\bar{\lambda}_{2}^{\alpha_{k}}+\bar{\lambda}_{3}^{\alpha_{k}}-3)
\label{Eq7}
\end{equation}
where $\bar{\lambda}_{k}$ are distortional principal stretches which can be calculated from principal stretches ($\lambda_{k}$) by $\bar{\lambda}_{k}=J^{-1/3}\lambda_{k}$, $N$ is the order of the strain energy potential ($N = 3$), $k$ is the index of summation ($k =1, 2, 3$), $J$ is the elastic volume strain, $J={\lambda}_1 {\lambda}_2 {\lambda}_3$, and $\mu_k$ and $\alpha_k$ are the material constants. Principal strains governed failure was employed to capture the fracture behavior of the elastomer \cite{Drass2019}
\begin{equation}
\varepsilon_{eq}=\underset{i=1,2}{max}\left \{\varepsilon_i \right\}
\label{Eq8}
\end{equation}
\begin{equation}
\varepsilon_1,\varepsilon_2=\frac{\varepsilon_{xx}+\varepsilon_{yy}}{2} \pm \sqrt{\left(\frac{\varepsilon_{xx}-\varepsilon_{yy}}{2}\right)^2+\left(\frac{\gamma_{xy}}{2}\right)}  
\label{Eq9}
\end{equation}
where $\varepsilon_{eq}$ is the equivalent strain. $\varepsilon_1$ and $\varepsilon_2$ are the principal strains. $\varepsilon_{yy}$ and $\varepsilon_{xx}$ are the engineering strain in loading and transverse directions, respectively. $\gamma_{xy}$ denotes the shear strain in the $x-y$ plane.

\begin{figure*}[t]
\centering
\includegraphics[width=0.95 \linewidth]{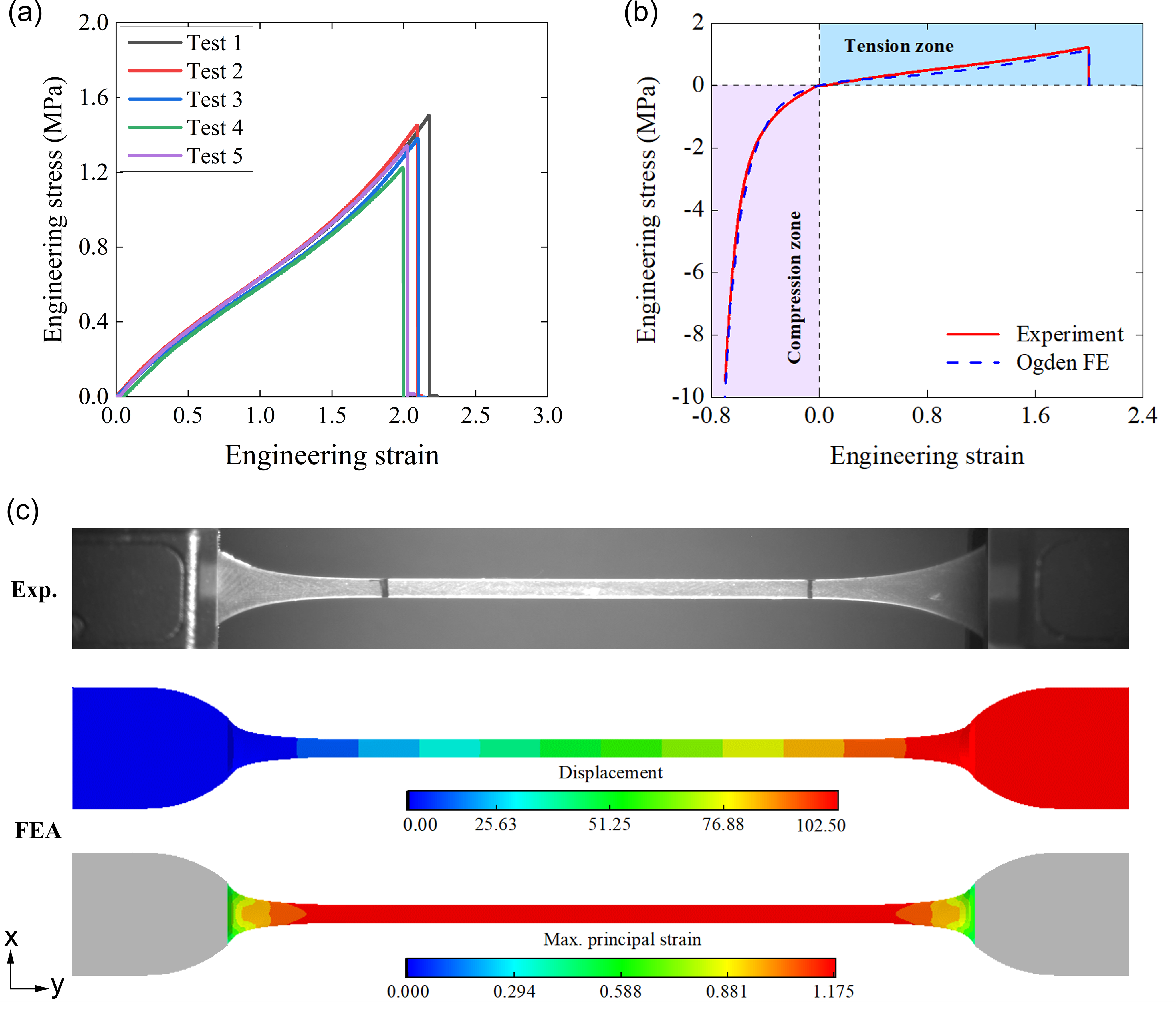}
\caption{(a) Five repetitive uniaxial tensile tests for the determination of average failure strain. (b) Comparison of the uniaxial mechanical behaviour between experiments and the Ogden model. (c) Maximum elongation captured from the experiment and the FE simulation. The maximum principal strain in the gauge section was used as the critical equivalent strain for failure prediction.}
\label{Fig3}
\end{figure*}

Uniaxial tensile tests for bulk elastomer were repeated five times, as shown in Fig. \ref{Fig3}a. Their average failure strain (1.175) in the gauge section was employed to determine the critical equivalent strain, as shown in Fig. \ref{Fig3}c. Figure \ref{Fig3}b shows that the uniaxial properties of the bulk elastomer can be accurately captured by Ogden's model using material data $\mu_1=-0.997$ MPa, $\mu_2=1.098$ MPa, $\mu_3=0.207$ MPa, $\alpha_1=2.000$, $\alpha_2=2.359$, and $\alpha_3=-1.724$.

\begin{figure*}[t]
\centering
\includegraphics[width=0.9 \linewidth]{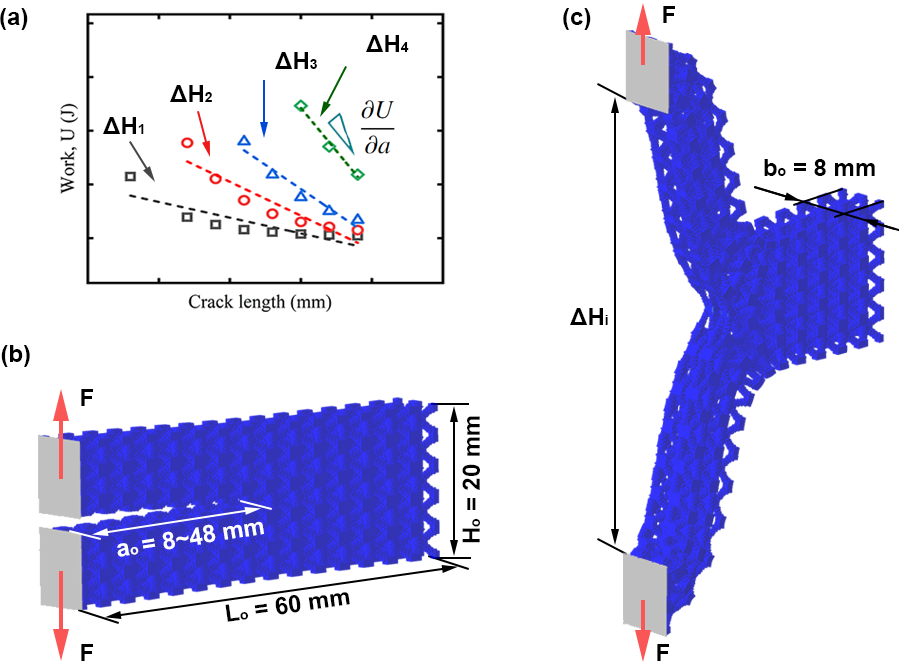}
\caption{(a) The work and crack lengths curves with different deformation levels to determine a suitable $\frac{\partial U}{\partial a}$. (b) and (c) dimensions of undeformed and deformed double peeling specimen for FE simulations, respectively.}
\label{Fig4}
\end{figure*}

\subsection{Validations to fracture energy measurements}
In addition to Eq.\ref{Eq3}, two other methods proposed by Rivlin and Thomas \cite{Rivlin1953} were implemented numerically to demonstrate the accuracy of the planar tension in the measurement of fracture energy. First, the planar tensile tests were simulated using notched samples with various initial crack sizes ($a = 8 \sim 48$), as shown Fig. \ref{Fig2}b. The force and displacement curves of notched specimens were obtained by loading the specimens to ultimate failure. The total energy $U$ stored in the notched samples at different levels of elongation ($\Delta H_o$) was calculated by the area under the force and displacement curve. The total energy $U$ of notched specimens was plotted with various crack lengths to determine the slope of the curve $\left (\frac{\partial U}{\partial a}\right)$, namely strain energy release rate per unit crack length, as shown Fig. \ref{Fig4}a. Then, fracture energy determined by notched specimens with various initial crack lengths is given by 
\begin{equation}
\Gamma =\frac{1}{b_o}\left(\frac{\partial U}{\partial a}\right)
\label{Eq10}
\end{equation}

Second, double peeling tests were also implemented numerically to further verify the accuracy of the fracture energy. Figures \ref{Fig4}b and c show the geometry of the undeformed and deformed double-peeling specimen, respectively. Samples with various initial crack sizes ($a = 8 \sim 48$) were studied to find a suitable $\left (\frac{\partial U}{\partial a}\right)$ to determine the fracture energy using Eq.\ref{Eq10}. In addition, the fracture energy of double peeling tests can be also determined by \cite{Gent1996,Tanaka2005}
\begin{equation}
\Gamma =\frac{2F}{b_o}
\label{Eq11}
\end{equation}

In the present study, the fracture energy determined by Eq. \ref{Eq3} was verified by above two methods, namely notched samples with various initial crack sizes using Eq. \ref{Eq10} and double peeling tests using Eq. \ref{Eq11}. For the former method, notched samples with various initial crack lengths can be used to either planar tension or double peeling test.

\subsection{Finite element modelling}
Full-scale and three-dimensional (3D) finite element models were developed to study the tensile and fracture mechanics of elastomer lattices using the commercial FE solver \emph{ABAQUS/Explicit}. The elastomer lattices were modelled with 8-node 3D linear brick elements with a reduced integration scheme (C3D8R elements). The properties of the materials were modeled using the incompressible Ogden model, as mentioned in Section \ref{S3.2}. Since the elastomer lattices are highly stretchable, truss structure may contact each other under large deformations. Therefore, general contact using frictionless tangential behaviour and 'hard' contact normal behaviour were defined to model the contact between the trusses. Both uniaxial and pure shear tests used the same boundary conditions, fixing a clapping end and applying a monotonically increased displacement at the free end. Mesh convergence tests were conducted to ensure that the results of the numerical simulations were independent of the mesh density. Various mesh sizes, ranging from a global mesh size of 0.1 to 0.8 $mm$, were tested, and the global mesh size of 0.2 $mm$ was selected for numerical simulations. To ensure that the simulations capture the quasi-static responses of the system, the loading rate was controlled to ensure that the kinetic energy of the system was within 5\% of the internal energy.

\section{Results and discussions}
\subsection{Tensile properties}
\begin{figure*}[!t]
\centering
\includegraphics[width=0.9 \linewidth]{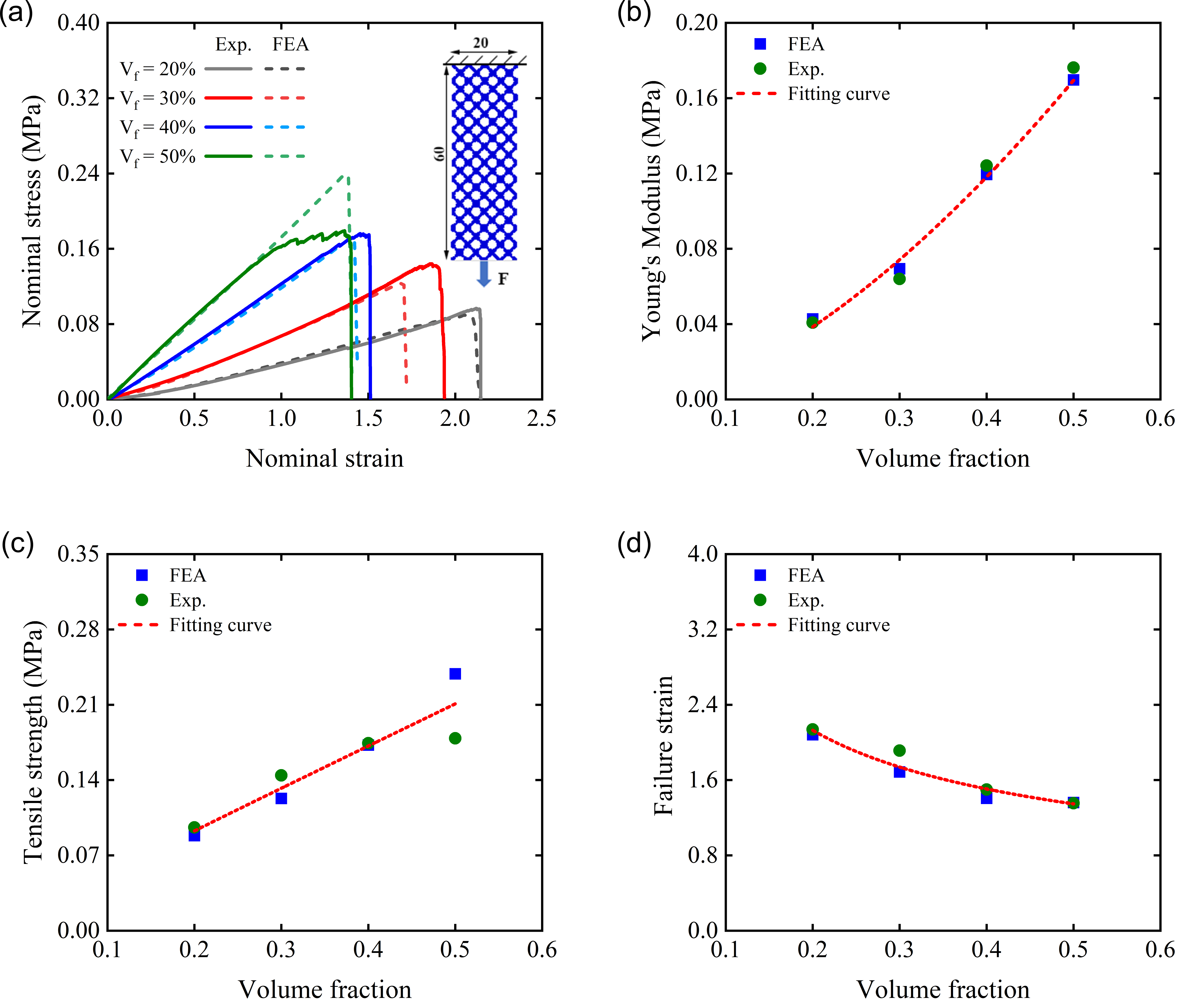}
\caption{(a) Uniaxial stress-strain curves of elastomer lattices with various volume fractions. (b) – (d) Variations of Young’s modulus, tensile strength, and failure strain in response to a volume fraction change.}
\label{Fig5}
\end{figure*}

Figure \ref{Fig5} shows the uniaxial tensile properties of the elastomer lattice with various volume fractions from 0.2 to 0.5. Figure \ref{Fig5}a shows a comparison of the stress-strain curves between experimental measurements and numerical simulation. The FE simulation was coincident with experiments throughout the loading process, which further validated the accuracy of the numerical models. Minor discrepancies were observed in tensile strength and failure strain, which were mainly attributed to manufacturing defects. These defects may lead to minor variations in the failure strain of individual trusses, which further leads to progressive failure of the lattice elastomer. However, in the FE simulation, each truss in an ordered lattice structure was assumed to have an identical failure strain that resulted in structure failure in a sudden manner. Figures \ref{Fig5}b-d show the tendency of Young's modulus, tensile strength, and failure strain with different volume fractions. Both FE simulations and experiments suggest that Young’s modulus and tensile strength increase as the volume fraction increases, whereas the failure strain decreases as volume fraction increases. These tensile properties (Figs. \ref{Fig5}b-d) can be fitted using Eq.\ref{Eq1} and the material constants in Table\ref{Tab1}.

\begin{table*}[!t]
\renewcommand\tabcolsep{3pt}
\small
\centering
\setlength{\abovecaptionskip}{5pt}
\caption{Summary of available instruments for the characterisation of wet adhesive behaviors.}
\label{Tab1}
\begin{tabular*}{1\textwidth}{c@{\extracolsep{\fill}} ccccccc}
\hlineB{2.5}\noalign{\medskip}
&$P_{s}$&&&&&$P_l$&\\
\cmidrule(lr){1-3}\cmidrule(lr){5-8}\noalign{\medskip}
       $E_s$&$\sigma_s$&$\varepsilon_s$& & &$E_l\propto E_s$&$\sigma_l\propto \sigma_s$&$\varepsilon_l\propto \varepsilon_s$ \\\noalign{\medskip}
         \multirow{2}{*}{0.61 MPa}& \multirow{2}{*}{1.21 MPa}&\multirow{2}{*}{2.00}&&$C$& 0.86 MPa & 0.32 MPa & 0.48\\\noalign{\medskip}
         &  &  &&$n$& 1.61 & 0.90 &-0.50\\\noalign{\medskip}
\hlineB{2.5}\noalign{\smallskip}
\end{tabular*}
\end{table*}

\begin{figure*}[!t]
\centering
\includegraphics[width=1 \linewidth]{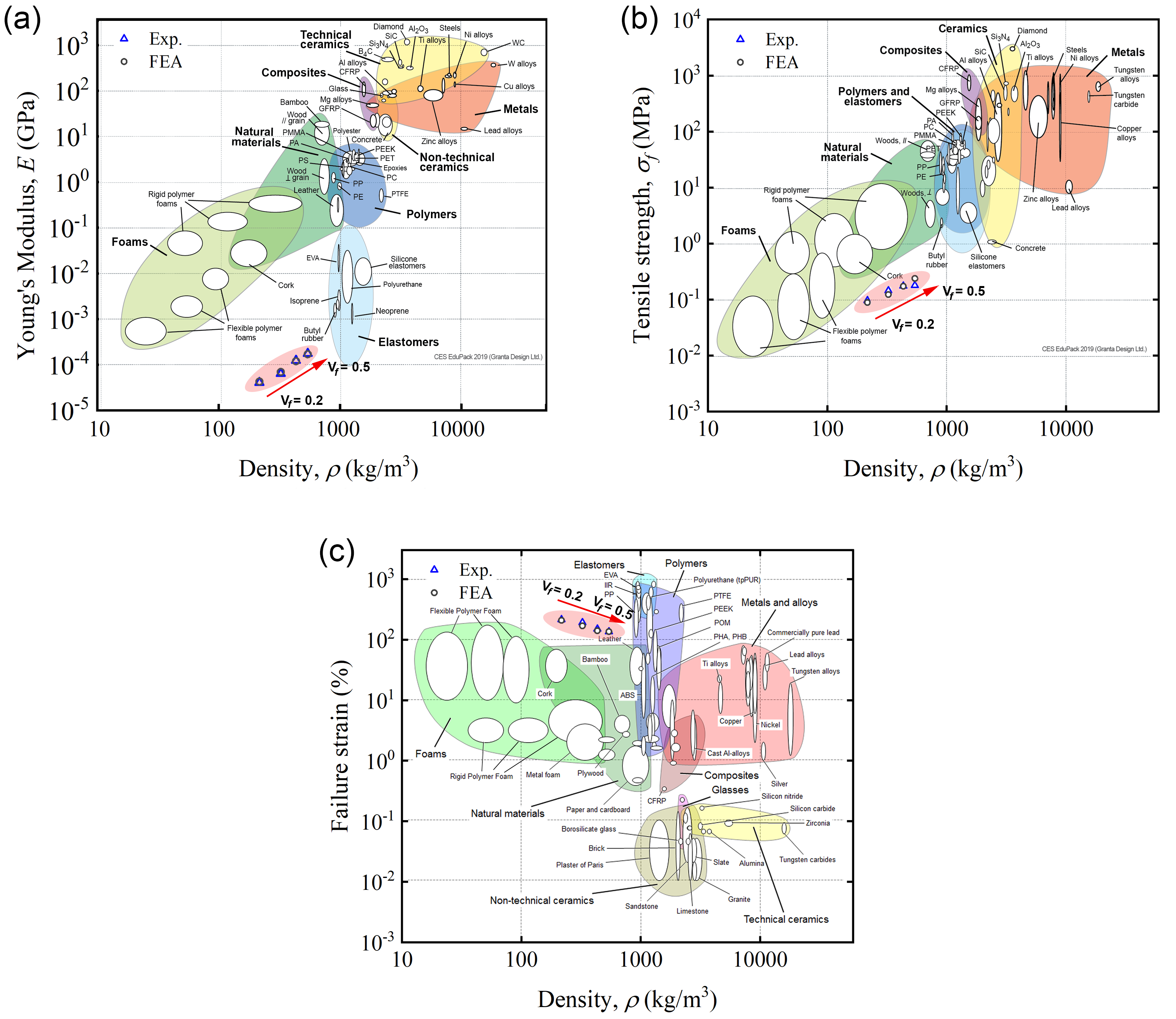}
\caption{Mechanical properties of elastomer lattices in Ashby diagram. (a) Young’s modulus vs. density map, (b) tensile strength vs. density map, and (c) failure strain vs. density map.}
\label{Fig6}
\end{figure*}

\begin{figure*}[!t]
\centering
\includegraphics[width=1 \linewidth]{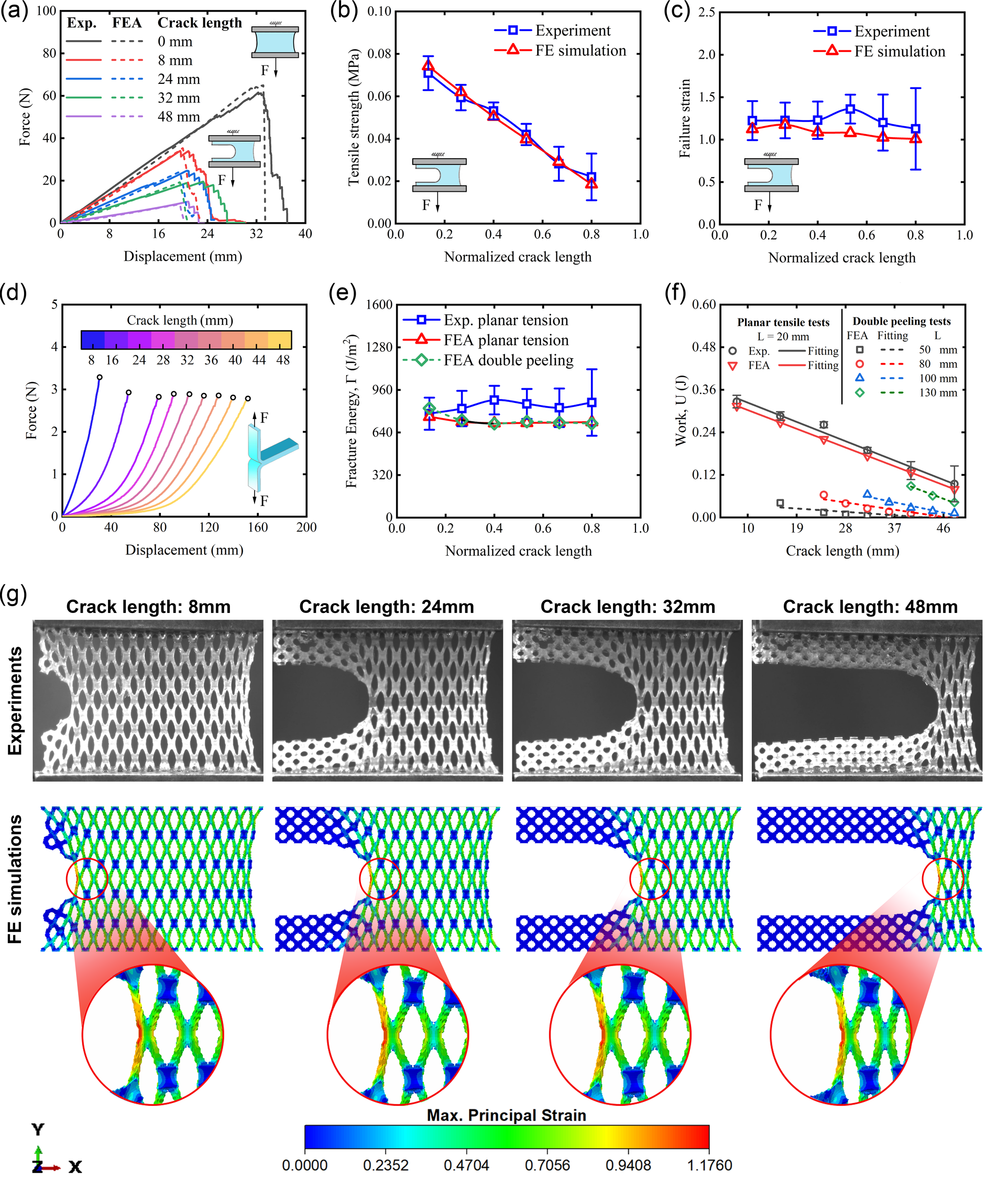}
\caption{(a) Representative force and displacement curves of planar tension for various crack lengths. (b) and (c) Failure strain and tensile strength of notched specimens as a function of the normalised crack length. (d) Force and displacement curves of double peeling tests for various crack lengths based on FE simulation. (e) Fracture energy of elastomer lattices for various normalized crack lengths. (f) Work vs. crack length curves for planar tensile and double peeling tests. (g) Comparisons of lattice deformation between FE simulations and experimental observations at the onset of crack propagation. Note: the experimental data are means ± SD (n = 3 for each crack length).}
\label{Fig7}
\end{figure*}

\begin{figure*}[!t]
\centering
\includegraphics[width=1 \linewidth]{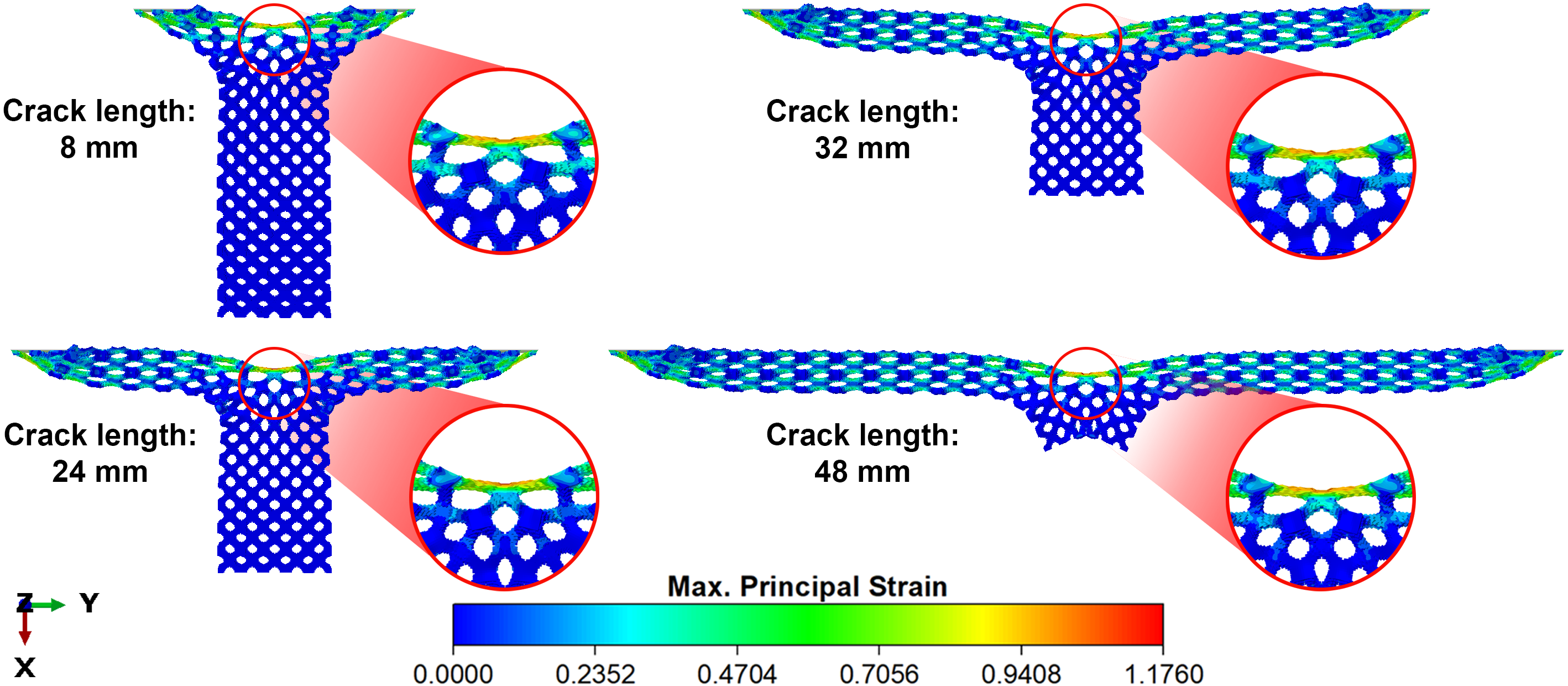}
\caption{FE simulated double peeling tests showing lattice deformations at the initiation of crack propagation for varying crack lengths}
\label{Fig7a}
\end{figure*}

The mechanical properties of elastomer lattices were compared with available materials in Ashby diagram. Figure \ref{Fig6}a shows that the elastomer lattices achieve the lowest Young's modulus in the Ashby diagram of approximately $0.04\sim0.17$ $MPa$ in a density range of $217\sim542$ $kg/m^3$. Such a low Young’s modulus suggests that the proposed elastomer lattices are softer than foams, bulk elastomers, polymers, and other engineering materials. Figure \ref{Fig6}b plots the tensile strength against the density map, which shows that the elastomer lattices achieve a moderate tensile strength ($0.096\sim0.179$ $MPa$) compared to the foams. However, Fig. \ref{Fig6}c shows that the proposed elastomer lattice achieves a higher failure strain ($135.5\% \sim 213.9\%$) compared to almost all foam materials. Notably, the failure strain of elastomer lattices almost approaching or even exceeding the bulk materials at low volume fraction ($\leq20\%$).

\subsection{Effects of crack length}
The effects of initial crack length on fracture energy were investigated using notched specimens with various normalised crack lengths ($\bar{a} = 0.13 \sim 0.80$). Figure \ref{Fig7}a shows the representative force and displacement curves of notched and un-notched specimens subjected to planar tension. The results show that the FE simulation coincided well with the experimental measurement. An increase in initial crack lengths resulted in a reduction in the stiffness of the notched specimen. Both the FE and the experimental results show that the elongation at the onset of crack propagation and the ultimate failure are insensitive to the initial crack length. Figure \ref{Fig7}b shows that the tensile strength decreases proportionally from 0.07 to 0.02 MPa as the crack length increases. Figure \ref{Fig7}c shows the failure strain of FE simulation and experiments as a function of the normalized crack length. Notably, failure strain remains consistent, independent of crack length, with all notched specimens failing at strains between 1.12 and 1.36. Figure \ref{Fig7}g shows comparisons of lattice deformation between FE simulations and experimental observations at the onset of crack propagation. The experimental results suggest that the deformations of the samples in which the crack begins are independent of the initial length of the crack. FE simulations show an identical distribution of maximum principal strain at crack tips, which further demonstrates that the crack depth would not alter the failure mechanism of elastomer lattices subjected to planer tension.

\begin{figure*}[!t]
\centering
\includegraphics[width=1 \linewidth]{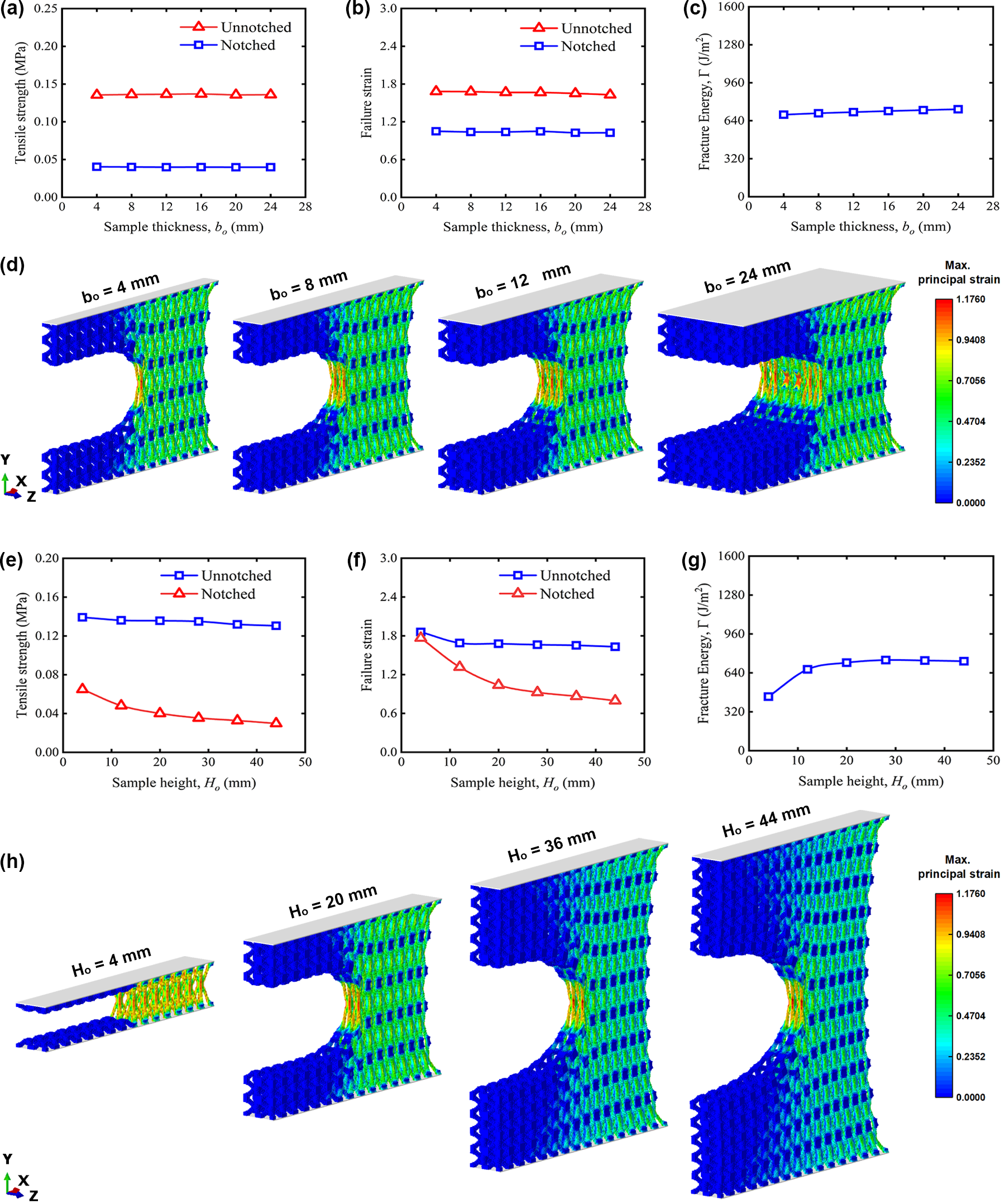}
\caption{(a) Tensile strength, (b) failure strain, and (c) fracture energy of elastomer lattice as a function of initial sample thickness ($b_o$). (e) FE simulated lattice deformations for various sample thicknesses at the onset of crack propagation. (a) Tensile strength, (b) failure strain, and (c) fracture energy of elastomer lattice as a function of initial sample height ($H_o$). (e) FE simulated lattice deformations for various sample heights at the onset of crack propagation.}
\label{Fig8}
\end{figure*}

Double peeling tests proposed by Rivlin and Thomas \cite{Rivlin1953} were simulated to verify the accuracy of the fracture energy based on the planar tension. Figure \ref{Fig7}d shows the force-displacement curves of the double peeling simulations. The results suggest that the critical force to initiate crack propagation remains constant with initial crack length ranging from $24 \sim 48$ $mm$. These critical forces were further employed to calculate the fracture energy using Eq. \ref{Eq11} and then plotted in Fig.\ref{Fig7}e. Figure \ref{Fig7}e shows that the fracture energy of the elastomer lattices does not vary significantly when the normalized crack length was greater than 0.27. Figure \ref{Fig8} shows lattice deformations at the onset of crack propagation for different crack lengths. It was observed that the deformation at the crack tip remains identical regardless of the varying crack lengths. The fracture energy based on FE simulations (planar tension and double peeling) is relatively higher at a normalized crack length of 0.13, approaching a stable value of approximately 710 $J/m^2$ when the normalized crack length exceeds 0.27. Although the fracture energies measured from experiments fluctuate with increasing crack length, they agreed well with the FE simulations. The discrepancies between the FE simulations and the experiments are less than 20\%, which is within the acceptable level ($\pm30\%$) recommended by Rivlin and Thomas \cite{Rivlin1953}. Furthermore, Fig. \ref{Fig7}f shows the total energy stored in deforming the specimen with various crack lengths. In the double peeling simulation, different levels of deformation ($L = 50$ to $130$ $mm$) were tested and a suitable deformation ($L = 130$ $mm$) was selected to calculate the slope $\left (\frac{\partial U}{\partial a}\right)$ of the work-crack length curve. In the planar tensile test, the deformation ($L = 20$ $mm$) at the onset of crack propagation was selected to determine $\frac{\partial U}{\partial a}$. The fracture energies obtained from this method are approximately 756 $J/m^2$ (planar tension) and 720 $J/m^2$ (double peeling test), which are highly consistent with the experimental measurements and FE simulations in Fig.\ref{Fig7}e. It should be noted that even though FE simulations and experiments show that the fracture energy is independent of normalized crack length, deep or shallow cracks introduce a large deviation of up to 250 $J/m^2$ in the experiment. Consequently, the normalized crack length within the range of 0.27 to 0.67 ensures consistent measurements in fracture energy. 

\subsection{Dimension dependency}
Planer tensions were simulated to investigate the effects of sample dimensions on fracture properties. The notched elastomer lattices were modeled with an initial width $L_o = 60$ $mm$, an initial height $H_o =20$ $mm$ and a crack length $a = 32$ $mm$. Various thicknesses ($b_o = 4 \sim 24$ $mm$) were used to investigate their effects on fracture properties. Figures \ref{Fig8}a and b show the tensile strength and failure strain of the elastomer lattices as a function of the thickness of the sample, respectively. The results suggest that the tensile strength and failure strain of the notched and unnotched specimens are insensitive to the thickness of the sample. Figure \ref{Fig8}c shows that the fracture energy remained nearly constant (around 720 $J/m^2$) as the thickness increased. Figure \ref{Fig8}d shows the maximum principal strain distribution at the initiation of crack propagation. The results show a consistent strain distribution at the tip of the crack among the notched lattices with thicknesses ranging from 4 to 12 $mm$. In contrast, due to compression induced by Poisson's effects, the 24 $mm$ thick elastomer lattice displays a higher deformation at the center than the free surface in the thickness direction. Although the central region in the thickness direction of the crack tip is more vulnerable to early failure, this does not significantly affect the measurement of the fracture energy.

\begin{figure*}[!t]
\centering
\includegraphics[width=1 \linewidth]{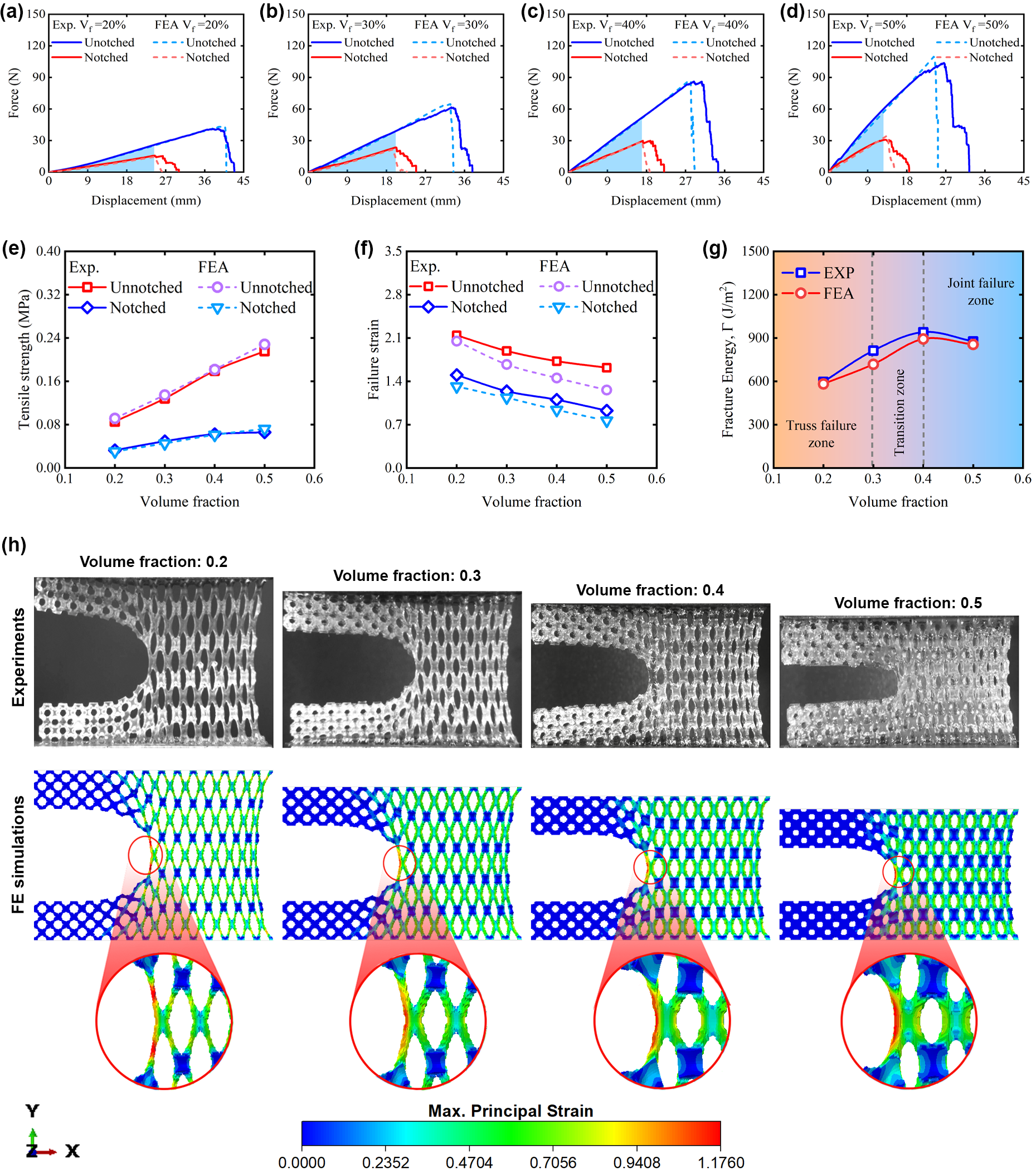}
\caption{(a) - (d) Representative force and displacement curves with various volume fractions. (e) Tensile strength and (f) failure strain of notched and un-notched specimens as a function of volume fractions. (g) fracture energy of elastomer lattice with various volume fractions. (h) A comparison of lattice deformation between FE simulation and experimental observations at the onset of crack propagation.}
\label{Fig10}
\end{figure*}

In addition, the notched specimens were modeled using various initial heights $H_o= 4 \sim 44$ $mm$ to investigate the effects of the sample height on the fracture properties. The initial width $L_o = 60$ $mm$, the thickness $b_o =8$ $mm$, and the crack length $a = 32$ $mm$ were fixed for these FE simulations. Figures \ref{Fig8}e and f show that the tensile strength and failure strain of the notched specimens decrease as the initial height increases. For the unnotched specimens, both tensile strength and failure strain tend to stabilize, reaching a constant value when the initial height exceeds 20 $mm$. Figure \ref{Fig8}g plots the fracture energy as a function of initial height, which shows that the fracture energy converges to a stable value of approximately 720 $J/m^2$ at an initial height of 20 $mm$. Figure \ref{Fig8}h shows the deformation of notched lattices with various initial heights when they are subjected to planer tension. It was found that the distribution of maximum principal strain at the crack tip becomes identical when the initial height is greater than 20 $mm$. Consequently, this specific height was adopted for the notched elastomer lattices in Section \ref{S4.4}.

\subsection{Effects of volume fraction}
\label{S4.4}
Notched and unnotched elastomer lattices were studied numerically and experimentally to investigate the effects of volume fraction on fracture energy. Figures \ref{Fig10}a-d show the representative force and displacement curves of the elastomer lattices subjected to planar tension. The numerical simulations were in good agreement with experimental tests. A larger volume fraction provides a higher stiffness (the slope of force and displacement curves) for both notched and un-notched lattices but reduces its elongation. Figures \ref{Fig10}e and f show the tensile strength and failure strain of the notched and unnotched specimens as a function of volume fractions. Similar to the uniaxial tension, tensile strength is proportional to the volume fraction, whereas the failure strain is inversely proportional to the volume fraction. Both FE simulation and experiments suggested that a crack of $\bar{a} = 0.47$ results in a significant reduction in tensile strength (up to 70\%) and failure strain (up to 42\%). 

\begin{figure*}[!t]
\centering
\includegraphics[width=0.65 \linewidth]{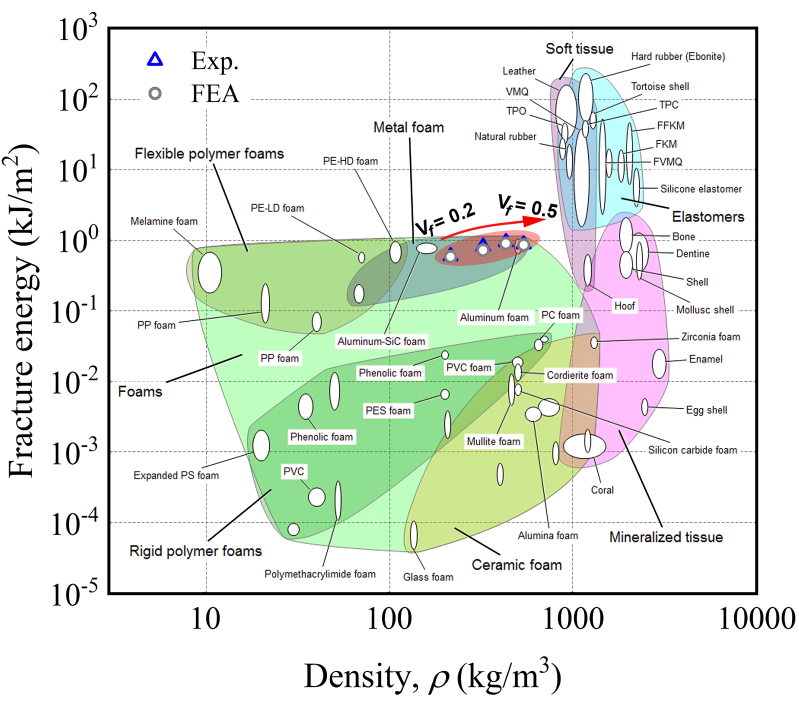}
\caption{Fracture energy vs. density map for the elastomer lattice with various volume fractions.}
\label{Fig11}
\end{figure*}

In addition, Fig. \ref{Fig10}g shows that the fracture energy of the elastomer lattice exhibits an upward trend till the 40\% volume fraction and a downward trend afterwards. To delve deeper into the failure mechanisms at distinct volume fractions, an exploration of lattice deformation at the onset of crack propagation was conducted through both FE simulations and experimental observations, as shown in Fig. \ref{Fig10}h. The results suggest that at a low volume fraction (e.g., 10\%), the lattice trusses at the crack tip exhibit a high degree of freedom (DoF) in in-plane rotation, forming a blunt crack tip where individual trusses undergo significant deformation against crack propagation. At moderate volume fractions (e.g. 30\% or 40\%), the deformation at single trusses diminishes, while the deformation at truss joints intensifies. At a high volume fraction (e.g. 50\%), the DoF of single trusses is further constrained due to the thicker truss joint. In this scenario, elastomer lattices form a sharper crack than at a low-volume fraction, resulting in a more substantial deformation at the truss joint and ultimately culminating in premature failure. Thus, Fig. \ref{Fig10}g can be divided into three zones, namely truss failure, transition, and joint failure zones, based on the distinct failure modes exhibited by lattices at varying volume fractions.

Figure \ref{Fig11} presents the fracture energy and density map for the elastomer lattices in different volume fractions. The results reveal that these elastomer lattices exhibit a fracture energy of approximately $582 \sim 941$ $J/m^2$ within a density range of $217 \sim 542$ $kg/m^3$. This notable fracture energy places them close to the upper limit of engineering foam materials, surpassing the fracture energy of many ceramic, flexible, and rigid polymer foams. Although the fracture energy is marginally lower than that of solid soft tissue and elastomers, the proposed elastomer lattices approach the lower boundary of these benchmarks while maintaining a significantly lower density.

\section{Conclusions}
Marine mussel plaque-inspired soft, ductile, and tough lattices have been successfully developed, manufactured, and tested. The main conclusions drawn from the present work are summarised as follows.

\begin{itemize}
\setlength\itemsep{0em}
\item[•] Quasi-static uniaxial tensile tests were performed on elastomer lattices with volume fractions from 0.2 to 0.5. Both the FE and experimental results indicate that Young’s modulus and tensile strength exhibit an upward trend, while failure strain experiences a downward trend with an increase in volume fraction.

\item[•] The applicability of the Rivlin and Thomas \cite{Rivlin1953} methods to determine the fracture energy of the elastomer lattices has been confirmed. The findings indicate that the fracture energy determined through planar tension consistently aligns with values obtained using two alternative approaches: tensile tests using samples with various crack lengths and double peeling tests.

\item[•] The effects of sample dimensions and volume fractions on the fracture energy of the proposed elastomer lattices were investigated. It was found that the fracture energy is insensitive to initial crack length, thickness, and height ($H_o\geq 20$ $mm$). Two distinct failure modes, truss failure and joint failure, were correlated with volume fraction. The elastomer lattice with a 40\% volume fraction was found to fail at the transition of these two failure modes and achieve the highest fracture energy.

\item[•] Inspired by marine mussel plaque, lattice structures manufactured by elastomers provide remarkable softness, ductility, and toughness. Through the manipulation of volume fraction, the proposed elastomer lattices achieved Young’s module of $0.04\sim0.17$ $MPa$, failure strain of $135.5\% \sim 213.9\%$, and fracture energy of $582 \sim 941$ $J/m^2$. These properties positioned the elastomer lattices as highly softer, more ductile, and tougher than most engineering foams in the Ashby diagram.

\end{itemize}

\newpage
\makenomenclature
\nomlabelwidth=30mm 
\nomenclature{\(\Gamma\)}{Fracture energy of the elastomer lattice.}
\nomenclature{\(b_o\), \(H_o\), \(L_o\)}{Initial sample thickness, height, and width.}
\nomenclature{\(\bar{a}\), \(a_o\)}{Normalised and initial crack length.}
\nomenclature{\(P_s\), \(P_l\)}{Mechanical properties of the bulk materials and elastomer lattices.}
\nomenclature{\(\rho_s\), \(\rho_l\)}{Densities of the bulk materials and elastomer lattices.} 
\nomenclature{\(\rho\)}{Density of engineering materials.} 
\nomenclature{\(V_f\)}{Volume fraction of the elastomer lattice.} 
\nomenclature{\(E_s\), \(E_l\)}{Young’s modulus of the bulk materials and elastomer lattices.} 
\nomenclature{\(\sigma_s\), \(\sigma_l\)}{Tensile strength of the bulk materials and elastomer lattices.} 
\nomenclature{\(\varepsilon_s\), \(\varepsilon_l\)}{Failure strain of the bulk materials and elastomer lattices.} 
\nomenclature{\(\Delta H_i\)}{Elongation of the elastomer lattices.}
\nomenclature{\(C\), \(n\)}{Coefficients used to estimate mechanical properties of lattices.}
\nomenclature{\(H_c\)}{Critical elongation of notched specimen at the onset of crack propagation.}
\nomenclature{\(\bar{\lambda}_{k}\), \(\lambda_{k}\)}{Distortional principal stretches and principal stretches of Ogden’s mode.}
\nomenclature{\(N\), \(k\)}{The order of the strain energy potential and index of summation of Ogden’s model.}
\nomenclature{\(\mu_k\), \(\alpha_k\)}{Material constants of Ogden’s model.}
\nomenclature{\(J\)}{Elastic volume strain.} 
\nomenclature{\(\varepsilon_{eq}\)}{Equivalent strain.} 
\nomenclature{\(\varepsilon_{1}\), \(\varepsilon_{2}\)}{Maximum and minimum principal strain.} 
\nomenclature{\(\varepsilon_{xx}\), \(\varepsilon_{yy}\)}{Engineering strain in transverse and longitudinal directions.} 
\nomenclature{\(\gamma_{xy}\)}{Shear strain.} 
\printnomenclature

\section*{Acknowledgement}
The authors acknowledge the use of instruments and scientific and technical assistance at the NanoVision Centre, Queen Mary University London.
\section*{Funding}
This work was funded by the Leverhulme Trust Research Grant Scheme, UK (No. RPG-2020-235) and EPSRC Research Grant (EP/X017559/1).
\section*{Competing interests}
The authors of the manuscript have no competing interests.

\section*{Author contributions}
Y. Pang: Writing - Manuscript preparation, Experimentation, Methodology, Software, Conceptualisation;  T. Liu: Conceptualization, Methodology, Writing - Review and Editing, Supervision, Funding Acquisition

\biboptions{numbers,sort&compress}
\bibliographystyle{elsarticle-num}
\bibliography{References} 

\begin{thebibliography}{10}
\expandafter\ifx\csname url\endcsname\relax
  \def\url#1{\texttt{#1}}\fi
\expandafter\ifx\csname urlprefix\endcsname\relax\def\urlprefix{URL }\fi
\expandafter\ifx\csname href\endcsname\relax
  \def\href#1#2{#2} \def\path#1{#1}\fi

\bibitem{Fernandes2021}
M.~C. Fernandes, J.~Aizenberg, J.~C. Weaver, K.~Bertoldi, Mechanically robust
  lattices inspired by deep-sea glass sponges, Nature Materials 20~(2) (2021)
  237--241.

\bibitem{Ha2022}
N.~San~Ha, T.~M. Pham, T.~T. Tran, H.~Hao, G.~Lu, Mechanical properties and
  energy absorption of bio-inspired hierarchical circular honeycomb, Composites
  Part B: Engineering 236 (2022) 109818.

\bibitem{Du2018}
A.~Du~Plessis, C.~Broeckhoven, I.~Yadroitsev, I.~Yadroitsava, S.~G. le~Roux,
  Analyzing nature's protective design: the glyptodont body armor, Journal of
  the Mechanical Behavior of Biomedical Materials 82 (2018) 218--223.

\bibitem{Jia2019}
Z.~Jia, Y.~Yu, S.~Hou, L.~Wang, Biomimetic architected materials with improved
  dynamic performance, Journal of the Mechanics and Physics of Solids 125
  (2019) 178--197.

\bibitem{Fu2019}
J.~Fu, Q.~Liu, K.~Liufu, Y.~Deng, J.~Fang, Q.~Li, Design of bionic-bamboo
  thin-walled structures for energy absorption, Thin-Walled Structures 135
  (2019) 400--413.

\bibitem{Zou2016}
M.~Zou, S.~Xu, C.~Wei, H.~Wang, Z.~Liu, A bionic method for the crashworthiness
  design of thin-walled structures inspired by bamboo, Thin-Walled Structures
  101 (2016) 222--230.

\bibitem{Tan2011}
T.~Tan, N.~Rahbar, S.~Allameh, S.~Kwofie, D.~Dissmore, K.~Ghavami, W.~Soboyejo,
  Mechanical properties of functionally graded hierarchical bamboo structures,
  Acta biomaterialia 7~(10) (2011) 3796--3803.

\bibitem{Wang2019}
H.~Wang, D.~Gu, K.~Lin, L.~Xi, L.~Yuan, Compressive properties of bio-inspired
  reticulated shell structures processed by selective laser melting, Advanced
  Engineering Materials 21~(4) (2019) 1801168.

\bibitem{Gu2017}
G.~X. Gu, M.~Takaffoli, M.~J. Buehler, Hierarchically enhanced impact
  resistance of bioinspired composites, Advanced Materials 29~(28) (2017)
  1700060.

\bibitem{Dimas2013}
L.~S. Dimas, G.~H. Bratzel, I.~Eylon, M.~J. Buehler, Tough composites inspired
  by mineralized natural materials: computation, 3d printing, and testing,
  Advanced Functional Materials 23~(36) (2013) 4629--4638.

\bibitem{Marthe2009}
M.~Rousseau, A.~Meibom, M.~G{\`e}ze, X.~Bourrat, M.~Angellier, E.~Lopez,
  Dynamics of sheet nacre formation in bivalves, Journal of structural biology
  165~(3) (2009) 190--195.

\bibitem{Qureshi2022}
D.~A. Qureshi, S.~Goffredo, Y.~Kim, Y.~Han, M.~Guo, S.~Ryu, Z.~Qin, Why mussel
  byssal plaques are tiny yet strong in attachment, Matter 5~(2) (2022)
  710--724.

\bibitem{Cohen2019}
N.~Cohen, J.~H. Waite, R.~M. McMeeking, M.~T. Valentine, Force distribution and
  multiscale mechanics in the mussel byssus, Philos. Trans. R. Soc. B: Biol.
  Sci. 374~(1784) (2019) 20190202.

\bibitem{Wilhelm2017}
M.~H. Wilhelm, E.~Filippidi, J.~H. Waite, M.~T. Valentine, Influence of
  multi-cycle loading on the structure and mechanics of marine mussel plaques,
  Soft Matter 13~(40) (2017) 7381--7388.

\bibitem{Desmond2015}
K.~W. Desmond, N.~A. Zacchia, J.~H. Waite, M.~T. Valentine, Dynamics of mussel
  plaque detachment, Soft matter 11~(34) (2015) 6832--6839.

\bibitem{Pang2023}
Y.~Pang, T.~Liu, Quasi-static responses of marine mussel plaques attached to
  deformable wet substrates under directional tensions, arXiv preprint
  arXiv:2305.15129 (2023).

\bibitem{Valois2019}
E.~Valois, C.~Hoffman, D.~G. Demartini, J.~H. Waite, The thiol-rich interlayer
  in the shell/core architecture of mussel byssal threads, Langmuir 35~(48)
  (2019) 15985--15991.

\bibitem{Jehle2020}
F.~Jehle, E.~Mac{\'\i}as-S{\'a}nchez, S.~Sviben, P.~Fratzl, L.~Bertinetti,
  M.~J. Harrington, Hierarchically-structured metalloprotein composite coatings
  biofabricated from co-existing condensed liquid phases, Nat. Commun. 11~(1)
  (2020) 1--9.

\bibitem{Harrington2010}
M.~J. Harrington, A.~Masic, N.~Holten-Andersen, J.~H. Waite, P.~Fratzl,
  Iron-clad fibers: a metal-based biological strategy for hard flexible
  coatings, Science 328~(5975) (2010) 216--220.

\bibitem{Filippidi2015}
E.~Filippidi, D.~G. DeMartini, P.~Malo~de Molina, E.~W. Danner, J.~Kim, M.~E.
  Helgeson, J.~H. Waite, M.~T. Valentine, The microscopic network structure of
  mussel (mytilus) adhesive plaques, J. R. Soc. Interface 12~(113) (2015)
  20150827.

\bibitem{Bernstein2020}
J.~H. Bernstein, E.~Filippidi, J.~H. Waite, M.~T. Valentine, Effects of sea
  water ph on marine mussel plaque maturation, Soft Matter 16~(40) (2020)
  9339--9346.

\bibitem{Bhuwal2023}
A.~S. Bhuwal, Y.~Pang, I.~Ashcroft, W.~Sun, T.~Liu, Discovery of
  quasi-disordered truss metamaterials inspired by natural cellular materials,
  J. Mech. Phys. Solids (2023) 105294.

\bibitem{Bhuwal2022}
A.~Bhuwal, Y.~Pang, T.~Liu, I.~Ashcroft, W.~Sun, Data-driven design of high
  ductile metamaterials under uniaxial tension, in: Current Perspectives and
  New Directions in Mechanics, Modelling and Design of Structural Systems, CRC
  Press, 2022, pp. 333--338.

\bibitem{Yang2022}
T.~Yang, H.~Chen, Z.~Jia, Z.~Deng, L.~Chen, E.~M. Peterman, J.~C. Weaver,
  L.~Li, A damage-tolerant, dual-scale, single-crystalline microlattice in the
  knobby starfish, protoreaster nodosus, Science 375~(6581) (2022) 647--652.

\bibitem{Pham2019}
M.-S. Pham, C.~Liu, I.~Todd, J.~Lertthanasarn, Damage-tolerant architected
  materials inspired by crystal microstructure, Nature 565~(7739) (2019)
  305--311.

\bibitem{Ge2021}
Q.~Ge, Z.~Chen, J.~Cheng, B.~Zhang, Y.-F. Zhang, H.~Li, X.~He, C.~Yuan, J.~Liu,
  S.~Magdassi, et~al., 3d printing of highly stretchable hydrogel with diverse
  uv curable polymers, Science advances 7~(2) (2021) eaba4261.

\bibitem{Luo2020}
C.~Luo, C.~Chung, N.~A. Traugutt, C.~M. Yakacki, K.~N. Long, K.~Yu, 3d printing
  of liquid crystal elastomer foams for enhanced energy dissipation under
  mechanical insult, ACS applied materials \& interfaces 13~(11) (2020)
  12698--12708.

\bibitem{Nicholas2020}
N.~A. Traugutt, D.~Mistry, C.~Luo, K.~Yu, Q.~Ge, C.~M. Yakacki,
  Liquid-crystal-elastomer-based dissipative structures by digital light
  processing 3d printing, Advanced Materials 32~(28) (2020) 2000797.

\bibitem{Yan2020}
D.~Yan, J.~Chang, H.~Zhang, J.~Liu, H.~Song, Z.~Xue, F.~Zhang, Y.~Zhang, Soft
  three-dimensional network materials with rational bio-mimetic designs, Nature
  communications 11~(1) (2020) 1180.

\bibitem{Li2017}
T.~Li, Y.~Jiang, K.~Yu, Q.~Wang, Stretchable 3d lattice conductors, Soft Matter
  13~(42) (2017) 7731--7739.

\bibitem{Zhang2022}
Y.~Zhang, K.~Yu, K.~H. Lee, K.~Li, H.~Du, Q.~Wang, Mechanics of stretchy
  elastomer lattices, Journal of the Mechanics and Physics of Solids 159 (2022)
  104782.

\bibitem{Rivlin1953}
R.~Rivlin, A.~G. Thomas, Rupture of rubber. i. characteristic energy for
  tearing, Journal of polymer science 10~(3) (1953) 291--318.

\bibitem{Sun2012}
J.-Y. Sun, X.~Zhao, W.~R. Illeperuma, O.~Chaudhuri, K.~H. Oh, D.~J. Mooney,
  J.~J. Vlassak, Z.~Suo, Highly stretchable and tough hydrogels, Nature
  489~(7414) (2012) 133--136.

\bibitem{Lee2019}
S.~Lee, M.~Pharr, Sideways and stable crack propagation in a silicone
  elastomer, Proceedings of the National Academy of Sciences 116~(19) (2019)
  9251--9256.

\bibitem{Wang20192}
Z.~Wang, C.~Xiang, X.~Yao, P.~Le~Floch, J.~Mendez, Z.~Suo, Stretchable
  materials of high toughness and low hysteresis, Proceedings of the National
  Academy of Sciences 116~(13) (2019) 5967--5972.

\bibitem{Sanoja2021}
G.~E. Sanoja, X.~P. Morelle, J.~Comtet, C.~J. Yeh, M.~Ciccotti, C.~Creton, Why
  is mechanical fatigue different from toughness in elastomers? the role of
  damage by polymer chain scission, Science advances 7~(42) (2021) eabg9410.

\bibitem{Liu2023}
B.~Liu, T.~Yin, J.~Zhu, D.~Zhao, H.~Yu, S.~Qu, W.~Yang, Tough and
  fatigue-resistant polymer networks by crack tip softening, Proceedings of the
  National Academy of Sciences 120~(6) (2023) e2217781120.

\bibitem{Lin2022}
S.~Lin, C.~D. Londono, D.~Zheng, X.~Zhao, An extreme toughening mechanism for
  soft materials, Soft Matter 18~(31) (2022) 5742--5749.

\bibitem{Shrimali2023}
B.~Shrimali, O.~Lopez-Pamies, The “pure-shear” fracture test for
  viscoelastic elastomers and its revelation on griffith fracture, Extreme
  Mechanics Letters 58 (2023) 101944.

\bibitem{Long2016}
R.~Long, C.-Y. Hui, Fracture toughness of hydrogels: measurement and
  interpretation, Soft Matter 12~(39) (2016) 8069--8086.

\bibitem{Maskery2022}
I.~Maskery, L.~Parry, D.~Padr{\~a}o, R.~Hague, I.~Ashcroft, Flatt pack: A
  research-focussed lattice design program, Additive Manufacturing 49 (2022)
  102510.

\bibitem{ASTMD412}
{ASTM D412-16}, Standard test methods for vulcanized rubber and thermoplastic
  elastomers-tension, ASTM International, West Conshohocken, PA. (2016).

\bibitem{Fleck2010}
N.~A. Fleck, V.~S. Deshpande, M.~F. Ashby, Micro-architectured materials: past,
  present and future, Proc R Soc Lond A Math Phys Sci. 466~(2121) (2010)
  2495--2516.

\bibitem{Gibson2003}
L.~J. Gibson, Cellular solids, Mrs Bulletin 28~(4) (2003) 270--274.

\bibitem{Liviu2020}
L.~Marșavina, E.~Linul, Fracture toughness of rigid polymeric foams: A review,
  Fatigue \& Fracture of Engineering Materials \& Structures 43~(11) (2020)
  2483--2514.

\bibitem{Anshul2023}
A.~Shrivastava, M.~Supreeth, N.~Gundiah, Crack propagation and arrests in
  gelatin hydrogels are linked to tip curvatures, Soft Matter 19~(36) (2023)
  6911--6919.

\bibitem{Ogden1972}
R.~W. Ogden, Large deformation isotropic elasticity--on the correlation of
  theory and experiment for incompressible rubberlike solids, Proc R Soc Lond A
  Math Phys Sci. 326~(1567) (1972) 565--584.

\bibitem{Drass2019}
P.~Rosendahl, M.~Drass, J.~Felger, J.~Schneider, W.~Becker, Equivalent strain
  failure criterion for multiaxially loaded incompressible hyperelastic
  elastomers, International Journal of Solids and Structures 166 (2019) 32--46.

\bibitem{Gent1996}
A.~Gent, Adhesion and strength of viscoelastic solids. is there a relationship
  between adhesion and bulk properties?, Langmuir 12~(19) (1996) 4492--4496.

\bibitem{Tanaka2005}
Y.~Tanaka, R.~Kuwabara, Y.-H. Na, T.~Kurokawa, J.~P. Gong, Y.~Osada,
  Determination of fracture energy of high strength double network hydrogels,
  The Journal of Physical Chemistry B 109~(23) (2005) 11559--11562.

\end{thebibliography}
\end{document}